\documentstyle[11pt,aaspp4,epsf,rotate]{article}

\def\psim{\lower.5ex\hbox{$\; \buildrel \propto \over \sim \;$}}
\begin{document}

\title{The Synchrotron Peak Shift during High-Energy Flares of Blazars}

\author{Markus B\"ottcher\altaffilmark{1}}

\altaffiltext{1}{Department of Space Physics and Astronomy, Rice University,
6100 Main Street, Houston, TX  77005-1892}

\centerline{\it To appear in Astrophysical Journal Letters}

\begin{abstract}

A prediction for the energy shift of the synchrotron
spectrum of flat-spectrum radio quasars (FSRQs) during 
high-energy flares is presented. If the $\gamma$-ray emission
of FSRQs is produced by Comptonization of external radiation,
then the peak of the synchrotron spectrum is predicted to 
move to lower energies in the flare state. This is opposite 
to the well-known broadband spectral behavior of high-frequency 
peaked BL Lac objects where the external radiation field is 
believed to be weak and synchrotron-self Compton scattering 
might be the dominant $\gamma$-ray radiation mechanism. The 
synchrotron peak shift, if observed in FSRQs, can thus be used 
as a diagnostic to determine the dominant radiation mechanism 
in these objects. I suggest a few FSRQs as promising candidates 
to test the prediction of the external-Comptonization model.

\end{abstract}


\section{Introduction}

66 blazar-type AGNs have been detected by the EGRET 
instrument on board the {\it Compton Gamma-Ray 
Observatory} as sources of $\gamma$-rays above 100~MeV  
(Hartman et al. \markcite{hart99}1999). These objects are  
identified with flat-spectrum radio sources classified as
BL Lac objects or flat-spectrum radio quasars (FSRQs). Many 
of these objects exhibit variability at all wavelengths, 
generally with the most rapid variability, on time scales of 
hours to days, observed at the highest $\gamma$-ray energies 
(e.g., Bloom et al. \markcite{bloom97}1997; Wagner et al. 
\markcite{wagner95}1995; Mukherjee et al. \markcite{muk97}1997).
The broadband (radio--$\gamma$-ray) emission from blazars is
most probably emitted via nonthermal synchrotron radiation and 
Comptonization of soft photons by energetic particles in relativistic 
outflows powered by accreting supermassive black holes. Potential 
sources of soft photons which are Compton-scattered to produce the 
$\gamma$-ray emission include internal synchrotron photons (e.g., 
Marscher \& Gear \markcite{mg85}1985, Maraschi et al. 
\markcite{maraschi92}1992, Bloom \& Marscher \markcite{bm96}1996), 
jet synchrotron radiation rescattered by circumnuclear material 
(Ghisellini \& Madau \markcite{gm96}1996, Bednarek 
\markcite{bed98}1998, B\"ottcher \& Dermer \markcite{bd98}1998),
and accretion-disk radiation which enters the jet directly (Dermer \& 
Schlickeiser \markcite{ds93}1993) and/or after being scattered by 
surrounding BLR clouds and circumnuclear debris (e.g., Sikora, 
Begelman \& Rees \markcite{sbr94}1994; Blandford \& Levinson 
\markcite{bl95}1995; Dermer, Sturner, \& Schlickeiser 
\markcite{dss97}1997; Protheroe \& Biermann \markcite{pb97}1997).

In recent years, extensive simultaneous broadband observations of 
blazars have enabled detailed modeling of their broadband 
spectra. A general result of these modeling efforts appears to
be that the spectra of high-frequency peaked BL Lac objects (HBLs) 
are well reproduced using a synchrotron-self Compton (SSC)
model, where the radio through X-ray radiation is produced by
nonthermal synchrotron radiation of ultrarelativistic electrons
and the $\gamma$-ray emission, extending in some cases up to
TeV energies, results from Compton scattering of the synchrotron
radiation by the same population of electrons (e. g., Mastichiadis
\& Kirk \markcite{mk97}1997 for Mrk~421, Pian et al.
\markcite{pian98}1998 for Mrk~501). However, a critical
question in this case is whether the primary acceleration
mechanism is able to accelerate electrons up to the required
ultrarelativistic energies, which are $\gg 1$~TeV if the
recent HEGRA detection of $> 25$~TeV photons from Mrk~501 
(Aharonian et al. \markcite{aha99}1999) is real. On the 
other hand, FSRQs are more successfully modeled assuming
that the soft seed photons for Comptonization are external 
to the jet (EC for external Comptonization; e. g., Dermer 
et al. \markcite{dss97}1997 for 3C~273, Sambruna et al. 
\markcite{sam97}1997 and Mukherjee et al. \markcite{muk99}1999 
for PKS~0528+134, B\"ottcher et al. \markcite{boe97}1997 for 3C~279). 

There appears to be a more or less continuous sequence 
of spectral properties of different objects, HBL $\to$ 
LBL $\to$ FSRQ, characterized by increasing $\gamma$-ray 
luminosity, increasing dominance of the energy output in the
$\gamma$-ray component over the synchrotron component, and a
shift of the peaks in the $\nu F_{\nu}$ spectra of both 
components towards lower energies. This sequence can be 
understood in terms of increasing dominance of the EC over 
the SSC mechanism (Fossati et al. \markcite{fos97}1997,
Ghisellini et al. \markcite{ghi98}1998). Recently, B\"ottcher
\& Collmar (\markcite{bc98}1998) and Mukherjee et al.
(\markcite{muk99}1999) have suggested that in the case
of FSRQs a similar sequence might also occur between 
different intensity states of the same object, and have 
applied this idea to the various states of PKS~0528+134.
There, it was assumed that the high states of PKS~0528+134
are characterized by a high bulk Lorentz factor of the
ultrarelativistic material in the jet, implying that the
external radiation field is more strongly boosted into
the comoving rest frame than during the quiescent state.
This effect, leading to a stronger dependence of the EC 
radiation on the Doppler factor than of the SSC radiation, 
was first pointed out by Dermer (\markcite{dermer95}1995).

In Fig. 1, the fit results to two extreme states of
PKS~0528+134 from Mukherjee et al. (\markcite{muk99}1999)
are compared. The figure reveals an important prediction
of the model adopted in that paper: During the $\gamma$-ray 
high state we expect that the synchrotron spectrum peaks at 
lower frequencies than in the low state. Unfortunately, the 
peaks of the $\nu F_{\nu}$ synchrotron spectra of most FSRQs 
are in the infrared and thus very hard to observe. For this
reason, the results of detailed modeling of the synchrotron 
component of an FSRQ have to be regarded with caution since 
in most cases the shape of the synchrotron spectrum is not 
well enough constrained to allow an exact determination
of jet parameters. Note, for example, that in Ghisellini
et al. (\markcite{ghi98}1998) in many cases the 9-parameter
model adopted there is used to effectively fit less than 
10 data points, and that the model IR -- optical spectra 
of PKS~0528+134 calculated in Mukherjee et al. 
(\markcite{1999}1999) for several observing periods are 
extremely poorly constrained. In the very low $\gamma$-ray
states of PKS~0528+134 (VPs 39, 337, and 616), even the 
$\gamma$-ray spectrum is very poorly constrained.

Therefore it is important to demonstrate that the predicted 
synchrotron peak shift does not depend on the details of the
adopted jet model and is not a consequence of fine-tuning
of parameters. In Section 2, I will present an analytical 
estimate of the synchrotron peak shift on the basis of very 
simple, general arguments. In Section 3, several FSRQs are 
suggested as promising candidates to test the prediction 
made in this Letter. I summarize in Section 4.

\section{Estimate of the synchrotron peak shift}

The shift of the $\nu F_{\nu}$ peak of the synchrotron spectrum
during a $\gamma$-ray flare of an FSRQ will be determined on the 
basis of the following general assumptions: (a) The $\gamma$-ray
flare is predominantly caused by an enhancement of the energy 
density $u'_{ext}$ of external photons in the rest frame comoving 
with a component (blob) of ultrarelativistic material moving along 
the jet. This enhancement can be caused by an increasing bulk 
Lorentz factor $\Gamma$ of the jet material. We have $u'_{ext}
= \Gamma^2 \, u_{ext}$.
(b) The electrons in the blob have a non-thermal distribution
with a peak at the energy $\gamma_b$ which is determined 
by the balance of an energy-independent acceleration rate 
$\dot\gamma_{acc}$ to the radiative energy loss rate (cf. 
Ghisellini et al. \markcite{ghi98}1998). 
(c) During a flare, the radiative energy loss rate of 
relativistic electrons in the blob is dominated by Compton 
scattering of external photons in the Thomson regime.

The assumptions (b) and (c) imply that $\gamma_b \propto 
\dot\gamma_{acc}^{1/2} \, u_{ext}^{-1/2} \, \Gamma^{-1}$. The 
jets of blazars are directed at a small angle $\theta \sim 
\Gamma^{-1}$ with respect to our line of sight. Thus, to a 
good approximation, the Doppler factor $\delta = \bigl(\Gamma 
\, [1 - \beta_{\Gamma} \cos\theta] \bigr)^{-1} \approx \Gamma$. 
Then, the observed peak of the synchrotron component depends on 
the bulk Lorentz factor and the external photon density as
\begin{equation}
\epsilon_{sy} \propto B' \, \gamma_b^2 \, \Gamma \propto
\dot\gamma_{acc} \, B' \, u_{ext}^{-1} \, \Gamma^{-1},
\end{equation}
where $B'$ is the magnetic field strength in the comoving frame. 
The apparent bolometric luminosity in the synchrotron component
varies acording to
\begin{equation}
L_{sy} \propto {B'}^2 \, \Gamma^4 \, \gamma_b^2 \propto 
\dot\gamma_{acc} \, {B'}^2 \, \Gamma^2 \, u_{ext}^{-1}.
\end{equation}
A plausible way to estimate the magnetic field strength might
be the assumption of equipartition of magnetic field energy
density to the energy density of ultrarelativistic electrons 
in the jet. However, our conclusions do not depend on the 
particular choice of $B'$. If equipartition applies, then 
$B' \propto \dot\gamma_{acc}^{1/4} \, u_{ext}^{-1/4} \, 
\Gamma^{-1/2}$, and Eqs.(1) and (2) become (the superscript 
$ep$ denotes the equipartition case)
\begin{equation}
\epsilon_{sy}^{ep} \propto \dot\gamma_{acc}^{5/4} \, 
u_{ext}^{-5/4} \, \Gamma^{-3/2}
\end{equation}
and 
\begin{equation}
L_{sy}^{ep} \propto \dot\gamma_{acc}^{3/2} \, u_{ext}^{-3/2} 
\, \Gamma.
\end{equation}
Assuming Thomson scattering at the $\nu F_{\nu}$ peak of the 
Compton component (this is a reasonable assumption for the
peak at several MeV -- 10~GeV, while beyond that energy 
Klein-Nishina effects may become important), the $\gamma$-ray 
spectrum peaks at
\begin{equation}
\epsilon_C \approx \epsilon_{ext} \, \gamma_b^2 \, \Gamma^2
\propto \epsilon_{ext} \, \dot\gamma_{acc} \, u_{ext}^{-1}
\end{equation}
where $\epsilon_{ext}$ is the mean photon energy of the external
photon field in the stationary frame of the AGN. The apparent
bolometric luminosity in the Compton component depends on 
$\Gamma$ and $u_{ext}$ through
\begin{equation}
L_C \propto u_{ext} \, \Gamma^6 \, \gamma_b^2 \propto 
\dot\gamma_{acc} \, \Gamma^4,
\end{equation}
independent of the external photon density. Remarkably, this
implies that an enhancement of the external photon density
$u_{ext}$, e. g., due to structural changes of the circumnuclear
material does not produce a $\gamma$-ray flare, although it
leads to spectral variability, shifting both spectral components
to lower frequencies.

Now let us consider the effect of a variation of the bulk 
Lorentz factor as a possible cause of a $\gamma$-ray flare.
Eqs. (1), (3), and (5) predict a shift of the synchrotron 
peak to lower frequencies, while the $\gamma$-ray peak should 
remain at basically the same photon energy. The ratio of the 
luminosities in both components varies according to $L_C / L_{sy} 
\propto \Gamma^2 \, u_{ext} \, {B'}^{-2}$ or $L_C / L_{sy}^{ep} 
\propto \Gamma^3 \, u_{ext}^{3/2} \, \dot\gamma_{acc}^{-1/2}$,
respectively. If $u_{ext}$ and $\dot\gamma_{acc}$ remain constant 
and equipartition applies, this yields a variation $L_C \propto 
(L_{sy}^{ep})^4$, thus predicting a much stronger relative 
variation of the two spectral components than predicted by 
the SSC model which was ruled out for 3C~279 for this reason 
by the observation of a variation of the $\gamma$-ray component 
by amplitudes greater than the square of the amplitudes of 
variation of the synchrotron component (Wehrle et al. 
\markcite{wehrle98}1998). 

The synchrotron peak is still shifted to lower energies even
if the increase of the bulk Lorentz factor is physically related
to a more efficient particle acceleration in the comoving frame, 
as long as during the flare the product $\dot\gamma_{acc} \, 
\Gamma^{-1}$ is lower than during the quiescent state. In this 
case, Eqs. (1), (3), and (5) predict that the $\gamma$-ray 
spectrum becomes spectrally harder, while at the same time 
the synchrotron peak shifts to lower energies. In fact, 
spectral hardening of the $\gamma$-ray spectrum during
flares is a common feature in EGRET-detected FSRQs (e. g., Collmar
et al. \markcite{col97}1997, Hartman et al. \markcite{hart96}1996,
Wehrle et al. \markcite{wehrle98}1998).

This flaring behavior is in qualitative contrast to the flares
observed in HBLs where a shift of both the synchrotron and the
Compton peaks to higher frequencies is observed (e. g., Catanese
et al. \markcite{cat97}1997, Pian et al. \markcite{pian98}1998,
Kataoka et al. \markcite{kat99}1999). These objects are believed 
to be qualitatively different from FSRQs because
(a) the broad-line regions surrounding the central engine
are weak or absent and the isotropic luminosity of the central
accretion disk is generally weaker than in quasars, implying that
the external soft photon field is much weaker than in the FSRQ
case and hence negligible compared to the intrinsic synchrotron 
radiation field, and (b) Compton scattering events near the 
$\nu F_{\nu}$ peak of the $\gamma$-ray spectrum most probably 
occur in the extreme Klein-Nishina regime, thus rendering Compton 
cooling rather inefficient. The flaring behavior of these sources 
seems to be dominated by an enhanced efficiency of electron 
acceleration in the jet which shifts both spectral components
towards higher photon energies (Mastichiadis \& Kirk 
\markcite{mk97}1997, Pian et al. \markcite{pian98}1998).

Therefore, I propose that the qualitatively different behavior
of the synchrotron peak during a $\gamma$-ray flare may serve
as a diagnostic to determine the dominant electron cooling and
radiation mechanism at $\gamma$-ray energies.

\section{Candidate sources}

As mentioned earlier, most of the bright and well-observed
FSRQs (e. g., PKS~0528+134: Mukherjee et al. \markcite{muk96}1996,
\markcite{muk99} or 3C~279: Wehrle et al. \markcite{wehrle98}1998)
have their synchrotron peak in the infrared where it is very hard 
to observe. Fig. 3 of Werhle et al. (\markcite{wehrle98}1998)
seems to indicate a shift of the synchrotron peak of 3C~279 to 
lower frequencies during the 1996 flare state. However, the peak
frequency range is not covered in any of the observing periods
presented there. Thus, the determination of the peak frequency
from the existing data on 3C~279 would necessarily be model 
dependent. 

A much more promising candidate to test the predictions
presented in this Letter is its ``sister'' source 3C~273
(von Montigny et al. \markcite{vM97}1997). This source 
appears to be ideal for such a study for several reasons. 
(1) It is a very bright radio and IR source, persistently
detectable from radio through IR and optical frequencies.
(2) During flares it is a strong EGRET source, allowing
an easy detection of a $\gamma$-ray flare even with the
degraded sensitivity of EGRET.
(3) The strong big blue bump yields a very exact 
determination of the luminosity ($\approx 3 \cdot 
10^{46}$~erg~s$^{-1}$) and spectrum of the underlying 
accretion disc of the AGN which facilitates the normalization 
of model calculations.

In Fig. 2, the simultaneous radio spectra of three epochs 
before, during, and after the prominent $\gamma$-ray flare of 
3C~273 in 1993 November (von Montigny et al. \markcite{vM97}1997)
are presented. A comparison of the flare-state radio spectrum 
(open circles connected by solid lines) to the pre-flare
and post-flare spectra seems to indicate a softening of the 
synchrotron spectrum during the flare, in perfect agreement 
with the expectation if the $\gamma$-rays are produced by the 
EC mechanism.

At frequencies above 100~GHz, the radio spectra shown in 
Fig. 2 are simultaneous to within 1 day. However, the
radio spectrum shortly (i. e. a few days) prior to the 
1993 $\gamma$-ray flare was not monitored simultaneously
with reasonable spectral coverage at $> 100$~GHz. The 
pre-flare period closest to the 1993 $\gamma$-ray flare 
for which such a simultaneous radio spectrum was 
available, was about 100 days before the flare. At that
time, 3C273 was not in the field of view of EGRET so that
we can not be sure that the source was in its quiescent
state. Furthermore, the broadband spectrum of 3C273 shown
in von Montigny et al. (\markcite{vM97}1997) and in Kubo et
al. (\markcite{kubo98}1998) suggests that the actual
synchrotron peak is located at higher frequencies,
$\nu_{sy} \sim 10^{13}$~Hz, which have not been monitored
in regular, short time intervals in the von Montigny et
al. (\markcite{vM97}1997) campaign.

Therefore, although this result seems very promising, 
it needs more solid confirmation by future broadband campaigns 
with good spectral and temporal coverage of the $\sim 100$~GHz 
-- $10^{14}$~Hz frequency range in order to allow a reliable, 
model-independent determination of the synchrotron peak 
of precisely simultaneous snapshot spectra in different 
$\gamma$-ray states of the source. 

In order to test the predictions presented here, it is particularly
important that the synchrotron spectrum {\it prior to} a $\gamma$-ray 
flare is known. An apparent shift of the synchrotron peak towards 
higher frequencies following a $\gamma$-ray flare is also predicted 
by alternative models, for example by the Marscher \& Gear 
(\markcite{mg85}1985) SSC model where a radio flare is delayed
with respect to an outburst at $\gamma$-rays due to the finite 
synchrotron cooling time of electrons in the jet. 

A few other FSRQs also emit a high flux at multi-GHz frequencies
and might therefore allow the precise determination of the
synchrotron peak and its shift during $\gamma$-ray flares
in future campaigns. In particular, PKS~0420-014 (Radecke
et al. \markcite{rad95}1995, von Montigny et al. \markcite{vM95}1995),
PKS~0521-365 (Pian et al. \markcite{pian96}1996), and 3C454.3 
(= PKS~2251+158; von Montigny et al. \markcite{vm95}1995)
have successfully been observed with near-complete frequency
coverage of their synchrotron peak, located at $\sim 10^{12}$
-- $10^{13}$~Hz. These sources may serve as test cases for the
predictions presented in this Letter.

\section{Summary}

On the basis of very general arguments I have shown that the
synchrotron $\nu F_{\nu}$ peak in the broadband spectra of
EGRET-detected FSRQs is expected to shift towards lower
frequencies during $\gamma$-ray flares, if Comptonization of
external radiation is the dominant electron cooling and
radiation mechanism at $\gamma$-ray energies. This behavior
is qualitatively different from most probably SSC-dominated
HBLs where both the synchrotron and the $\gamma$-ray 
component shift towards higher frequencies during 
the flare state. I propose this qualitative difference
as a new diagnostic tool to distinguish between the two
competing radiation mechanisms potentially responsible for
the production of $\gamma$-rays in the jets of blazars.
Results of a broadband campaign on 3C~273 have been used
to support the prediction of the external-Comptonization
model, but future campaigns with more continuous frequency
and temporal coverage in the radio -- IR spectral range are 
needed to draw solid conclusions. Other promising candidates
for such campaigns include the FSRQs PKS~0420-014, PKS~0521-365,
and 3C454.3.

\acknowledgements I thank R. Mukherjee and P. Sreekumar for 
valuable discussions, drawing my attention to this problem,
and for useful comments on the manuscript. I also thank the
referee, S. D. Bloom, for very useful suggestions. This work 
was supported by NASA grant NAG~5-4055.

\eject

\begin{figure}
\epsfysize=12.5cm
\rotate[r]{
\epsffile[150 0 580 550]{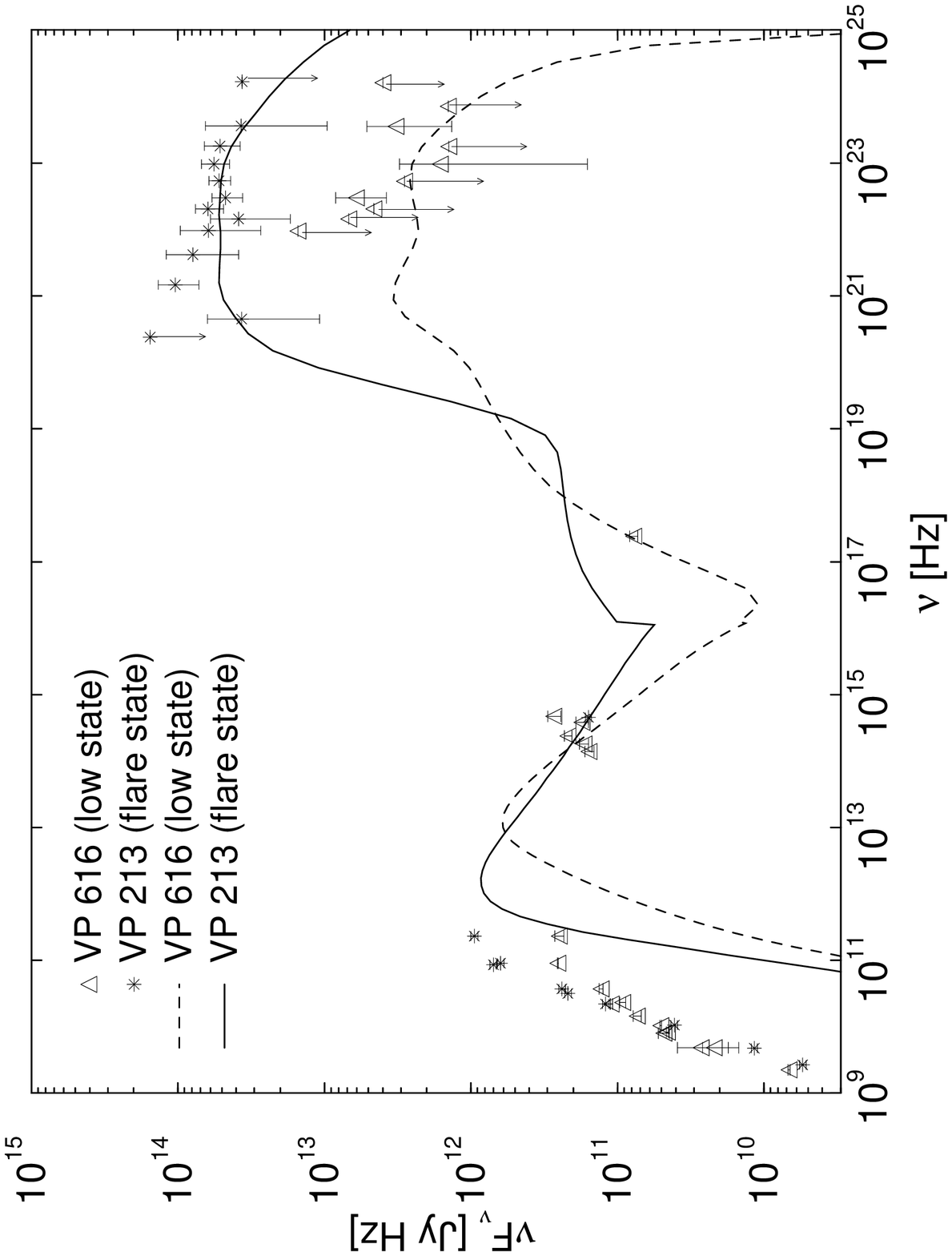}}
\caption[]{Fits to two broadband spectra of PKS~0528+134 in its
low and its flaring state, from Mukherjee et al. (\markcite{muk99}1999).
The predicted synchrotron peak shift to lower energies in the flaring
state is obvious.}
\end{figure}

\eject

\begin{figure}
\epsfysize=11cm
\rotate[r]{
\epsffile[200 50 550 500]{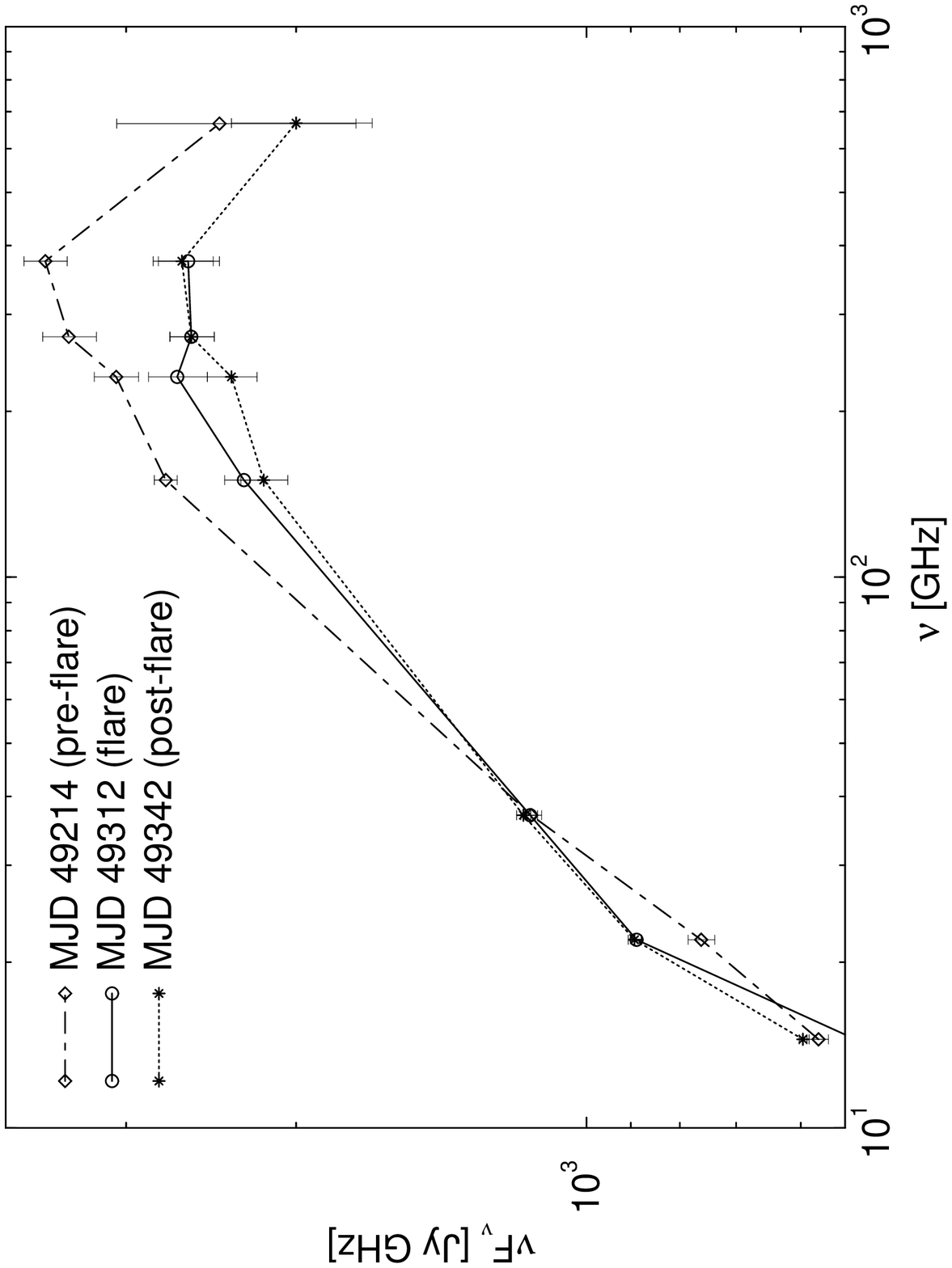}}
\caption[]{Evolution of the radio spectrum around the Nov.
1993 $\gamma$-ray flare of 3C~273. Data from von Montigny et al. 
(\markcite{vM97}1997). The figure indicates a softening of the 
radio spectrum during the $\gamma$-ray flare compared to the 
pre-flare and post-flare states.}
\end{figure}


\begin{references}

\reference{aha99} Aharonian, F., Akhperjanian, A. G., Barrio, J. A.,
et al., 1999, A\&A, 342, 69

\reference{bed98} Bednarek, W., 1998, A\&A, 336, 123

\reference{bl95} Blandford, R. D., \& Levinson, A. 1995, ApJ, 441, 79

\reference{bm96} Bloom, S. D., \& Marscher, A. P., 1996, ApJ, 461, 657

\reference{bloom97} Bloom, S. D., Thompson, D. J., Hartman, R. C., \& von
Montigny, C., 1997, in 4th Compton Symposium,  ed. C. D. Dermer, M. S.
Strickman, \& J. D. Kurfess (New York: AIP), 1262

\reference{bd98} B\"ottcher, M., \& Dermer, C. D., 1998, ApJ, 501, L51

\reference{boe97} B\"ottcher, M., Mause, H., \& Schlickeiser, R., 
Proc. of the 4th Compton Symposium, eds. C. D. Dermer, M. S. Strickman,
\& J. D. Kurfess, 1473

\reference{cat97} Catanese, M., Bradbury, S. M., Breslin, A. C.,
et al., 1997, ApJ, 487, L143

\reference{col97} Collmar, W., Bennett, K., Bloemen, H., et al., 1997,
A\&A, 328, 33

\reference{dermer95} Dermer, C. D., 1995, ApJ, 446, L63

\reference{ds93} Dermer, C. D., \& Schlickeiser, R., 1993, ApJ, 416, 458

\reference{dss97} Dermer, C. D., Sturner, S. J., \& Schlickeiser, R., ApJS, 109,
103

\reference{fos97} Fossati, G., Celotti, A., Ghisellini, G., \&
Maraschi, L., 1997, MNRAS, 289, 136

\reference{ghi98} Ghisellini, G., Celotti, A., Fossati, G., et al.,
1998, MNRAS, 301, 451

\reference{gm96} Ghisellini, G., \& Madau, P., 1996, MNRAS 280, 67

\reference{hart96} Hartman, R. C., Webb, J. R., Marscher, A. P.,
et al. 1996, ApJ, 461, 698

\reference{hart99} Hartman, R. C., et al., 1999, ApJ, in press

\reference{kat99} Kataoka, J., Mattox, J. R., Quinn, J., 1999,
ApJ, 513, in press

\reference{kubo98} Kubo, H., Takahashi, T., Madejski, G., et al.,
1998, ApJ, 504, 693

\reference{maraschi92} Maraschi, L., Ghisellini, G., \& Celotti, A., 1992, ApJ,
397, L5

\reference{mg85} Marscher, A. P., \& Gear, W. K., 1985, ApJ, 298, 114

\reference{mk97} Mastichiadis, A., \& Kirk, J. G., 1997, A\&A, 320, 19

\reference{muk96} Mukherjee, R., Dingus, B. L., Gear, W. K., et al.,
1996, ApJ, 470, 831

\reference{muk97} Mukherjee, R., Bertsch, D. L.,  Bloom, S. D.,
et al., 1997, ApJ 490, 116

\reference{muk99} Mukherjee, R., B\"ottcher, M., Hartman, R. C.,
et al., 1999, ApJ, submitted

\reference{pian96} Pian, E., Falomo, R., Ghisellini, G., et al.,
1996, ApJ, 459, 169

\reference{pian98} Pian, E., Vacanti, G., Gianpiero, T., et al., 
1998, ApJ, 492, L17

\reference{pb97} Protheroe, R. J., \& Biermann, P. L. 1997, Astroparticle 
Physics, 6, 293

\reference{rad95} Radecke, H.-D., Bertsch, D. L., Dingus, B. L.,
et al., 1995, ApJ, 438, 659

\reference{sam97} Sambruna, R. M., Urry, C. M., Maraschi, L.,
et al., 1997, ApJ, 474, 639

\reference{sbr94} Sikora, M., Begelman, M. C., \& Rees, M. J., 1994,  
ApJ, 421, 153

\reference{vM95} von Montigny, C., et al., 1995, ApJ, 440, 525

\reference{vM97} von Montigny, C., et al., 1997, ApJ, 483, 161

\reference{wagner95} Wagner, S. J., et al. 1995, ApJ, 454, L97

\reference{wehrle98} Wehrle, A. E., Pian, E., Urry, C. M., et al.\ 
1998, ApJ, 497, 178

\end{references}
\end{document}